\pdfoutput=1
\documentclass{article}
\usepackage{arxiv}

\usepackage{upgreek}
\usepackage{textcomp}
\usepackage{bm} 
\usepackage[utf8]{inputenc}
\usepackage[T1]{fontenc}
\usepackage{lmodern}

\usepackage{graphicx}
\usepackage[intlimits]{amsmath}
\usepackage{amssymb}
\usepackage{exscale}
\usepackage{hyperref}
\usepackage{times}
\usepackage{subfigure}
\usepackage{subfiles}
\usepackage{wrapfig}
\usepackage{floatrow}

\usepackage[square,sort,comma,numbers]{natbib}
\title{Nanocrystalline FeCr alloys synthesised by severe plastic deformation - a potential material for exchange bias and enhanced magnetostriction}

\author{
  Lukas Weissitsch \\
  Erich Schmid Institute of Materials Science, Austrian Academy of Sciences\\
  Jahnstra{\ss}e 12, 8700 Leoben, Austria \\
  \texttt{lukas.weissitsch@oeaw.ac.at}
\And
   Stefan Wurster \\
     Erich Schmid Institute of Materials Science, Austrian Academy of Sciences\\
  Jahnstra{\ss}e 12, 8700 Leoben, Austria\\ 
\And
     Alexander Paulischin\\
  Erich Schmid Institute of Materials Science, Austrian Academy of Sciences\\
  Jahnstra{\ss}e 12, 8700 Leoben, Austria \\
\And
     Martin St\"uckler \\
  Erich Schmid Institute of Materials Science, Austrian Academy of Sciences\\
  Jahnstra{\ss}e 12, 8700 Leoben, Austria \\
\And
  Reinhard Pippan\\
    Erich Schmid Institute of Materials Science, Austrian Academy of Sciences\\
  Jahnstra{\ss}e 12, 8700 Leoben, Austria 
\And
  Andrea Bachmaier\\
  Erich Schmid Institute of Materials Science, Austrian Academy of Sciences\\
  Jahnstra{\ss}e 12, 8700 Leoben, Austria 
}

\begin{document}
\maketitle

\begin{abstract}

This work gives insights into processing and characterisation of bulk nanocrystalline FeCr materials. The investigated FeCr alloys, consisting of 30, 50 and 70 at.\% ferromagnetic Fe and remaining anti-ferromagnetic Cr, are processed by arc melting and subsequent severe plastic deformation by high pressure torsion.
The physical similarities between elemental Fe and Cr in combination with the nanocrystalline structure of the as-deformed alloys, necessitates advanced characterisation techniques for the as-deformed state: In-situ annealing synchrotron X-ray diffraction measurements as well as electron microscopy experiments are linked to magnetostrictive measurements and reveal a single phase microstructure. 
Surprisingly, the nanocrystalline FeCr alloys remain supersaturated solid solutions upon annealing above 500~$^{\circ}$C, meaning a decomposition in a FeCr nanocomposite is suppressed. For the chosen annealing conditions grain growth is faster than decomposition and enhanced magnetostrictive values are found compared to materials in the as-deformed state.
\end{abstract}

 
\section{Introduction}
Tuning material properties by forming single phase alloys and varying chemical composition is often impeded by the existence of large miscibility gaps at the thermodynamic equilibrium of binary phase diagrams. 
Notwithstanding the difficulties in fabrication, there is a strong interest in such metastable phases created materials as they offer the opportunity to tune and enhance physical properties.
For quite some time, it is known to overcome these processing limits by techniques such as vapour deposition \cite{ma2005alloys}, mechanical alloying (MA) \cite{koyano1993mechanical} or severe plastic deformation (SPD) generating metastable phases \cite{weissitsch2021,bachmaier2012formation, kormout2017deformation}. 
The formation of supersaturated solid solutions by MA or SPD is accompanied by a high defect density and decreased microstructural sizes. This evinces characterisation challenges for the binary nanocrystalline FeCr system: Due to the same crystal structure and similar lattice parameters of elemental Fe and Cr, it is difficult to distinguish whether finely dispersed elemental nanocrystalline grains or a (supersaturated) solid solution is present \cite{koyano1993mechanical}. 
Possible characterisation methods are magnetic measurements.
Apart from the tuneability of the saturation magnetostriction of heterostructures \cite{szymczak1999mechanisms,han2017tailoring} Bormio-Nunes et. al \cite{bormio2016magnetostriction} reported an increasing saturation magnetostriction value for FeCr solid solutions. The aforementioned heterostructure of coexisting ferromagnetic Fe and anti-ferromagnetic Cr phases could give rise to an effect referred to as exchange bias.

The exchange bias features a shifted hysteresis loop, which originates from an unidirectional anisotropy created at the interface of the two types of magnetic materials.
This occurs, when the ferro-antiferromagnetic phase boundary is cooled through the Néel temperature (T$_{N}$) while a magnetic field is applied \cite{nogues1999exchange}.
Since magnetic exchange interactions appear on the nanometre-scale and the exchange bias effect is related to the phase boundaries, this effect is typically observed in microstructures obtained by well controllable synthetization techniques. Thin films or core-shell nanostructures \cite{binns2013exchange} are well controllable during processing and therefore, such material systems allow a proper investigation of fundamental physical properties \cite{giri2011exchange}. On the other hand, it is difficult to investigate the same effects in bulk materials \cite{mcdonald2019exchange}, exhibiting dimensions of several millimetres or larger. 
The bottleneck nowadays is the synthesis of a nanostructured bulk material exhibiting a high homogeneity therefore, enabling an up-scaling of materials with improved magnetic properties. 
Notwithstanding the above, the amount of interfaces is thought to directly influence the intensity of the magnetic effect, which makes magnetic nanostructured bulk materials necessary \cite{nogues1999exchange,chung2005interplay}.

Severe plastic deformation by high-pressure torsion (HPT) is a promising approach to overcome processing limitations, for above mentioned miscibility gaps of binary material systems \cite{kormout2017deformation} and furthermore a method of choice to tailor magnetic properties \cite{bachmaier2017tailoring, weissitsch2020}.
One big advantage of materials processed by HPT is the sample size, as they exhibit bulk dimensions up to several centimetres \cite{hohenwarter2019sample, horita2020severe}.
The principal goal of this work is the production of a bulk material consisting of alternating interfaces of elemental ferromagnetic Fe and anti-ferromagnetic Cr.
Therefore, the synthesis of a nanocrystalline supersaturated solid solution of FeCr by HPT is aimed for a subsequent annealing treatment. 
The increased total grain boundary area of nanocrystalline materials is expected to favour a decomposition \cite{SCHUH2015258} of the FeCr solid solution in separated phases and therefore, into a nanostructured composite.
A following magnetic field cooling is meant to induce an exchange bias, as already reported for HPT-compacted Co-NiO heterostructures \cite{menendez2007cold}.
However, processing and the characterization of nanocrystalline HPT deformed FeCr-alloys prior to field cooling is challenging. Evading processing limitations due to a pronounced miscibility gap in the binary phase diagram \cite{xiong2010phase}, we already reported an enormous increase in hardness during HPT-deformation but also on the successful processing and formation of a homogeneous nanocrystalline microstructure for a wide range of chemical compositions \cite{weissitsch2021}. 
In this work we clarify, if these materials exhibit a supersaturated solid solution or if they are already in a nanocrystalline two phase state. We report on in-situ high-energy X-ray diffraction (HEXRD) experiments where a solid solution without any decomposition upon annealing is presumed and verify these results by magnetostrictive measurements.

\section{Experimental}
Binary alloys consisting of nominal 30, 50 and 70 at.\% Fe and 70, 50, and 30 at.\% Cr, respectively, were processed by arc melting and subsequent HPT-deformation.
As starting materials, conventional flakes (Fe: 99.99+\% $<$~10~mm, Cr: 99.995\% 1-25~mm, both from HMW Hauner GmbH \& Co. KG) were used. All materials were stored and handled in an Ar-filled glovebox to protect the high purity materials against contamination. 

For arc-melting (AM/0.5 device, Edmund Buehler GmbH), a Ti-gettered high purity Ar atmosphere was used. To ensure chemical homogeneity the ingots were turned around and remelted at least five times. From arc melted ingots, cylinders with 8 mm diameter were prepared by electrical discharge machining and cut in discs with 1 mm thickness, which were subsequently deformed by HPT. 
A detailed description of the used HPT setup can be found in Ref. \cite{hohenwarter2009}.
A nominal pressure of 7.5~GPa with a rotational frequency of 0.6~min$^{-1}$ was used for the deformation process and 21 to 25 revolutions are applied (corresponding to a shear strain $\upgamma$ of $\sim$450 at radius \textit{r} = 3~mm \cite{valiev1999nanostructured}).
To avoid an increase of the processing temperature during deformation, high pressure air cooling was conducted.

After deformation, the sample is cut for further investigations and all measurements were performed at a radius $r\geqslant$~2~mm. Respective orientations are explained in detail in Reference \cite{weissitsch2020}. Scanning electron microscopy (SEM; LEO 1525, Carl Zeiss Microscopy GmbH) images were made in tangential HPT-disc orientation in backscattered electron (BSE) detection mode. 
The chemical composition was determined by energy dispersive X-ray spectroscopy (EDX; XFlash 6$\mid$60 device, Bruker), using the Bruker software package Esprit 2.2. The grain size of selected samples was quantitatively evaluated by Transmission Kikuchi Diffraction (TKD) and electron backscatter diffraction (EBSD) measurements using a Bruker e$^{-}$-Flash\textsuperscript{FS} detector.  
Hardness measurements (Micromet 5104, Buehler) were performed in tangential direction on the polished sample along the diameter in steps of $\Delta~r$ = 0.25 mm and were averaged for a radius 2~mm~$\leqslant~r\leqslant$~3.5~mm.
For annealing experiments, quarters or half discs of as-deformed samples were exposed to 300 $^{\circ}$C, 400 $^{\circ}$C and 500 $^{\circ}$C for 1~h in a conventional furnace. The samples were wrapped in a protective foil before heat treatment and were quenched with ethanol after the annealing process.

For magnetostrictive measurements, an electromagnet (Type B-E 30, Bruker) with conical poles (diameter 176~mm), providing a constant air gap (50~mm) and a maximal field of 2.25~T, was used. 
Within the air gap, the sample was placed into an encapsulating sample chamber, protecting from rapid temperature changes. The magnetostriction was measured using strain gauges (HBM, 1-LY11-0.6/120) applied with glue Z70 from HBM.
As a reference a piece of W, a very low magnetostrictive material is measured at the same time with a strain gauge of the same batch. According to \cite{kapitza1932study}, the magnetostriction of W can be expected to be below 0.005 ppm in a magnetic field of 2~T.
The strain gauge on the sample and the reference were placed parallel and on top of each other within the sample chamber and conducted to a half-Wheatstone bridge. The signal of the strain gauges was processed with an amplifier (HBM-Quantum\textsuperscript{x} MX410). Additionally, the constancy of the temperature was confirmed by a temperature measurement within the sample holder and the applied magnetic field was recorded, using a Hall-probe (Model 475 DSP, Lakeshore).

Synchrotron high-energy X-ray diffraction (HEXRD) experiments were carried out at the beamline P21.2 at PETRA III (DESY, Hamburg, Germany). The specimens were measured in transmission mode, while the spot size was set to 200 x 200~$\upmu$m$^{2}$ and a photon energy of 60 keV was used. 
In-situ annealing HEXRD experiments were performed using the following sequence:
Heating to 520~$^{\circ}$C, holding the temperature for 1 h and cooling. A Varex XRD 4343 flat panel detector captured diffraction patterns every 5 seconds. 
The experimental setup was realised by a heating microscope stage (THMS600 Linkam, Tadworth, United Kingdom), which was adapted to place the samples inside an Ar-flushed chamber to suppress the formation of oxides. The custom made chamber is designed to minimize the interaction of the beam with the furnace and allowed the diffracted beam to exit the probe chamber while the temperature was measured via a thermocouple. The temperature was calibrated by measuring a Cu sample with the same settings and comparing the changing lattice parameter during thermal expansion.
For data processing, the pyFAI package \cite{ashiotis2015fast} and further analysing the Matlab voigt package \cite{ruzi2020voigt} was used.

\section{Results}
\subsection{Characterisation of the microstructure}\label{sec:microstructure}
\begin{figure*}[h]
\centering
\includegraphics[width=\textwidth]{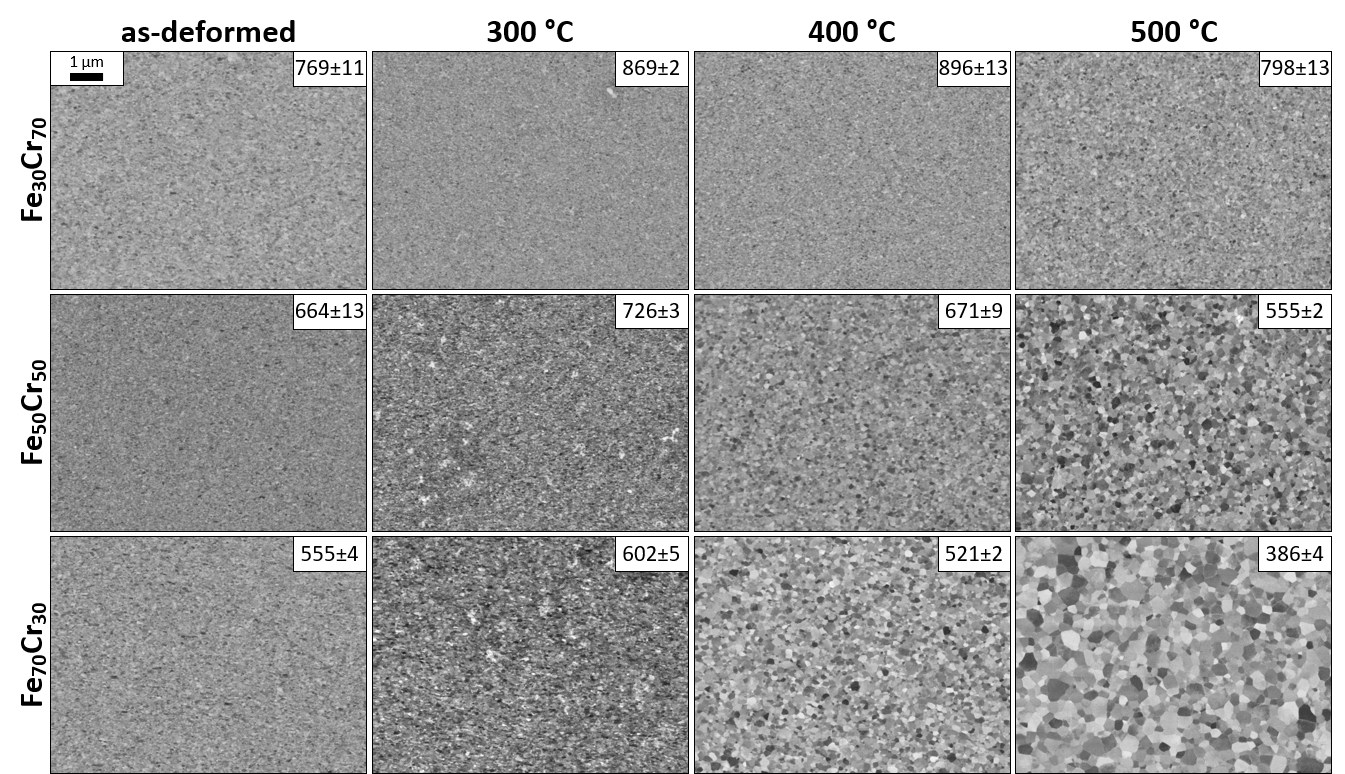}
\caption{BSE images of the microstructure of HPT-deformed FeCr alloys. The Fe content increases from the top to the bottom row. The first column shows the microstructure of as-deformed samples. The next columns depict microstructural changes with increasing annealing temperatures (300~$^{\circ}$C, 400~$^{\circ}$C and 500~$^{\circ}$C, respectively). Averaged Vickers-hardness values are depicted in the corresponding images as insets. The magnification is the same in all images.}
  \label{fig:microstructure}
\end{figure*}
\begin{figure}
\centering
\includegraphics[width=\columnwidth]{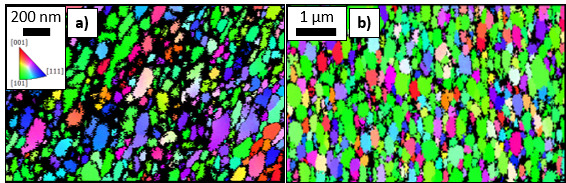}
\caption{Crystal orientation maps of \textbf{a)} an as-deformed Fe$_{50}$Cr$_{50}$ alloy recorded in axial HPT disc direction by TKD. \textbf{b)} EBSD scan of an Fe$_{50}$Cr$_{50}$ alloy after annealing at 500~$^{\circ}$C (1~h) measured in tangential direction but showing the crystal orientation in axial direction. Note the different scale bars.}
  \label{fig:TKD}
\end{figure}

The microstructures of room temperature (RT) HPT-deformed samples at r = 3~mm with different chemical compositions are shown in Figure \ref{fig:microstructure}. The Fe-content increases from the top to bottom row.
In the first column, BSE images of as-deformed FeCr alloys are depicted. Nanocrystalline and homogeneously deformed microstructures are visible. To quantify the grain sizes of the alloys in the as-deformed state, TKD is conducted exemplarily for one composition (Fe$_{50}$Cr$_{50}$). In Figure \ref{fig:TKD} a) a crystal orientation map revealed by TKD is shown. Recording several images and interpolating the data of 1256 grains with a log-normal distribution reveal a median grain size of 101~nm (standard deviation = 86~nm; $\sigma$ = 0.579, $\mu$ = -2.288). Please note the different scale bar in Figure \ref{fig:TKD} b), where an EBSD scan of the same alloy after annealing for 1h at 500~$^{\circ}$C is shown. The obtained median grain size is 307~nm (1456 grains, standard deviation = 160~nm; $\sigma$ = 0.449, $\mu$ = -1.180).
Results from Vickers-hardness testing are indicated as insets within each SEM micrograph (Figure \ref{fig:microstructure}). The hardness measurements along the radius, display a constant hardness, which indicates a homogeneously deformed state (denoted as 'as-deformed' in the following). To induce decomposition, the as-deformed samples are annealed for one hour at different temperatures (Figure \ref{fig:microstructure}). 
The change in the microstructure strongly depends on the chemical composition. Nearly no change in the SEM images for the Cr-rich sample (Fe$_{30}$Cr$_{70}$), depicted in the first row of Figure \ref{fig:microstructure}, is visible, when annealing up to 400~$^{\circ}$C. However, the hardness increases, which is a common feature for nanocrystalline materials \cite{renk2015increasing, huang2006hardening, valiev2010origin}, is covered. A slightly increased grain size, accompanied by a hardness decrease, is seen for the 500~$^{\circ}$C annealed sample.
The medium composition sample (Fe$_{50}$Cr$_{50}$) shows a slight contrast difference in SEM images between the as-deformed and 300~$^{\circ}$C annealed state. The microstructure remains almost constant even when annealing at 400~$^{\circ}$C (shown in the 3$^{rd}$ column) until a significant grain growth is found for 500~$^{\circ}$C annealing. The hardness increases upon annealing at 300~$^{\circ}$C but decreases for higher annealing temperatures. 
Differences for every state are visible for the Fe rich sample (Fe$_{70}$Cr$_{30}$). At 300~$^{\circ}$C, a slight grain growth is observed together with an increased hardness. Samples annealed at 400~$^{\circ}$C and 500~$^{\circ}$C exhibit a significant increase in grain size as well as a decrease of microhardness below the as-deformed state. 
Summarized, at intermediate temperatures the hardness increases for all chemical compositions, whereby the hardness peaks at higher temperatures for higher Cr content. Further, microstructural changes and grain growth are stronger pronounced for FeCr alloys containing a higher amount of Fe.
\begin{figure}
\centering
\includegraphics[width=\columnwidth]{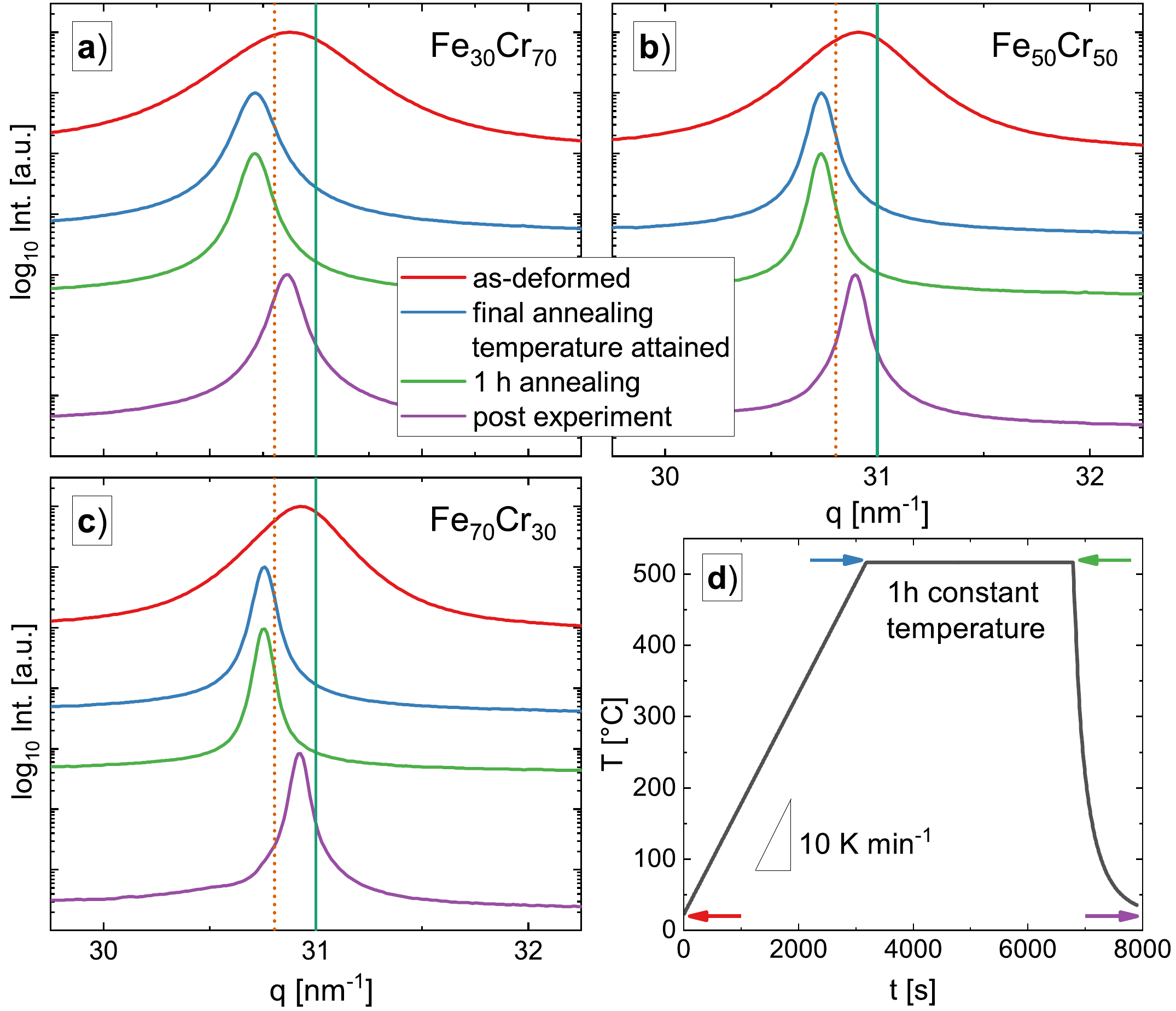}
\caption{HEXRD synchrotron data of the [110] peak for HPT-deformed FeCr samples consisting of \textbf{a)} Fe$_{30}$Cr$_{70}$, \textbf{b)} Fe$_{50}$Cr$_{50}$ and \textbf{c)} Fe$_{70}$Cr$_{30}$. Vertical lines denote theoretical reference lines for Fe (solid, turquoise) and Cr (dashed, brown). The selected patterns belong to in-situ annealing experiments and are extracted at different times and temperatures, visualized \textbf{d)} by arrows in an exemplarily temperature profile.}
  \label{fig:DESY_pattern}
\end{figure}
Laboratory XRD measurements of the as-deformed and annealed states (not shown) revealed broad peaks. They can either result from the formation of a supersaturated state or be due to the nanocrystalline state of individual Fe and Cr phases present in the as-deformed condition. 
Additionally, a co-existence of both configurations is conceivable.
Thus, to clarify this point and to further investigate the decomposition of FeCr alloys, which are possibly in a non-equilibrium state, in-situ annealing experiments using synchrotron XRD are performed. 
Due to the high brilliance of the beamline \cite{hegedus2019imaging}, it was expected to resolve individual peaks for Fe or Cr phases, if existent, and to precisely determine the phases of the as-deformed and annealed states.
Representatively, the most intense peak [110] of selected patterns of the in-situ experiment is shown in Figure \ref{fig:DESY_pattern} for all three FeCr compositions. The vertical lines indicate reference values for elemental Fe (solid, turquoise) and Cr (dashed, brown) calculated with lattice constants from \cite{landoltbornstein1995}.
The depicted peaks correspond to different temperatures during the in-situ experiment. 
The top peaks (red) belong to the beginning of the experiment, representing the as-deformed state. A significant peak broadening is again found for all three chemical compositions. 
According to Vegard's law \cite{vegard1921konstitution} the peak positions follow the trend of the chemical composition determined by EDX measurements of the FeCr alloys. The Fe contents obtained with EDX for the alloys depicted in Figure \ref{fig:DESY_pattern} a), b) and c) are 27.8$\pm$3.3 at.\%, 48.7$\pm$0.8 at.\% and 69.1$\pm$1.1 at.\%, respectively.
The blue peaks are recorded at the end of the heating process, when a constant temperature of 520~$^{\circ}$C is reached.
Because of the thermal expansion of the lattice, a shift towards a lower reciprocal lattice vector is visible. This is accompanied with a significantly sharpened shape of the peak, indicating larger crystallite sizes and/or diminishing crystallographic defects. 
The green peak is recorded after 1 hour of isothermal annealing at 520~$^{\circ}$C and shows almost no deviations with respect to the blue pattern. The purple peaks at the bottom correspond to an ex-situ measurement after the sample cooled down to room temperature and can be directly compared with the red line.
\begin{figure}
\centering
\includegraphics[width=\columnwidth]{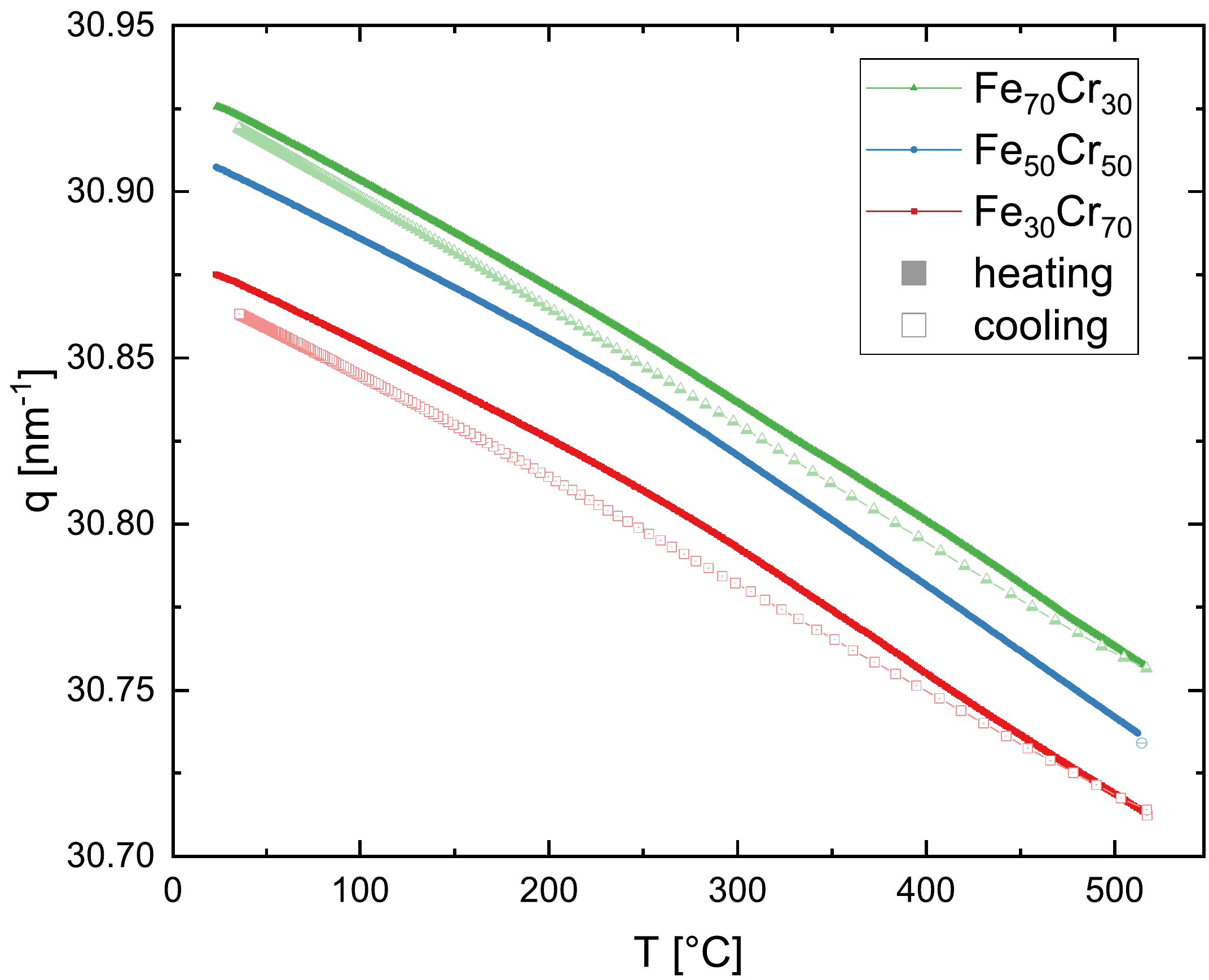}
\caption{[110] peak position for every pattern during heating (full symbols) and cooling (empty symbols) for HPT-deformed FeCr alloys determined by fitting the HEXRD in-situ data by a single Voigt profile. The cooling is not covered experimentally for the Fe$_{50}$Cr$_{50}$ alloy.}
  \label{fig:DESY_peakpos}
\end{figure}
The recorded patterns are fitted with Voigt-profiles. 
It should be mentioned, that for every peak one single Voigt-profile is used. 
Although fitting by one peak implies a solid solution, an overlapping of two peaks corresponding to separated Fe and Cr is also possible. Unfortunately, the similar crystal structures (both body centred cubic) and the similar lattice parameters with less than 0.02~\r{A} difference (see Reference \cite{landoltbornstein1995}) lead to very similar peaks for Fe and Cr, with almost no differences. Despite the high brilliance of the synchrotron, a pattern fitting with two separated Voigt-peaks is ambiguous. The shape, as well as intensity and position, strongly depend on the fitting starting parameters which allows no clear statement. 
Nevertheless, using a single Voigt peak profile, a high accordance is reached for the calculated pattern with respect to the measured HEXRD data. 
The peak positions as a function of the temperature process while heating (filled symbols) and cooling (empty symbols) is shown in Figure \ref{fig:DESY_peakpos}. A difference between these processes is recognised: While the peak position changes as expected linearly during cooling, the heating curve is bended. Although, no peak separation is found (see Figure \ref{fig:DESY_pattern}) the varying slope indicates a superposition of processes, quite likely dedicated to a change in the microstructure.

Therefore, the Voigt-profiles are analysed in more detail, as it is a convolution of a Gauss and Cauchy (Lorentz) distribution. 
The full width half maximum (FWHM) of a single Voigt profile can be parsed into the breadths of its Cauchy and Gaussian components. Keijser et al. \cite{keijser1982voigt} linked the individual peak widths to size (Cauchy) and strain (Gauss) broadening, respectively. The evolution of peak widths for each composition is shown in the supplementary data file. 
It is found that the Cauchy peak width decreases with increasing annealing temperature. In contrast, the Gaussian FWHM barely changes until 250~$^{\circ}$C before it decays rapidly with increasing temperature, indicating a reduction in internal stresses. 
The contribution of internal stresses to the overall peak width observed in Figure \ref{fig:DESY_pattern} is strongly reduced, when the heating process is finished. The remaining narrowing of the peaks, upon further isothermal annealing, originates from size effects and therefore grain growth, which is in accordance with SEM investigations of the annealed FeCr alloys (see Figure \ref{fig:microstructure}). 
After annealing for one hour at 520~$^{\circ}$C, no peak separation or broadening of the peak due to a superposition of its elemental Fe and Cr peaks is found. We therefore conclude, that all FeCr alloys exhibit a nanocrystalline, supersaturated solid solution state before and after the annealing treatment. To decompose these alloys, higher annealing temperatures are necessary.

\subsection{Magnetostrictive measurements}
In principle, 3d-elements like Fe exhibit a low magnetostrictive response \cite{james2000calculated} but when alloyed with Al, Si, Ge or Ga an increased value was found \cite{groessinger_2014}. This is explained by the substitution of an Fe atom by a non-magnetic metal.
The magnetostriction of arc melted FeCr alloys without subsequent plastic deformation was measured by Bormio-Nunes et al. \cite{bormio2016magnetostriction}. They processed FeCr solid solutions, with a Cr content between 5.3 and 30.7 at.\% and found an increased saturation magnetostriction compared to the magnetostriction of the pure elements. Besides the interesting question, if a higher Cr content further increases the magnetostriction, an enhanced value confirms our results from Section \ref{sec:microstructure}, as the increased magnetostriction is only present in the FeCr solid solution. 
Therefore, we conducted magnetostrictive measurements on all FeCr alloys in the as-deformed state as well as in one annealed state. 

\begin{figure}
\centering
\includegraphics[width=\columnwidth]{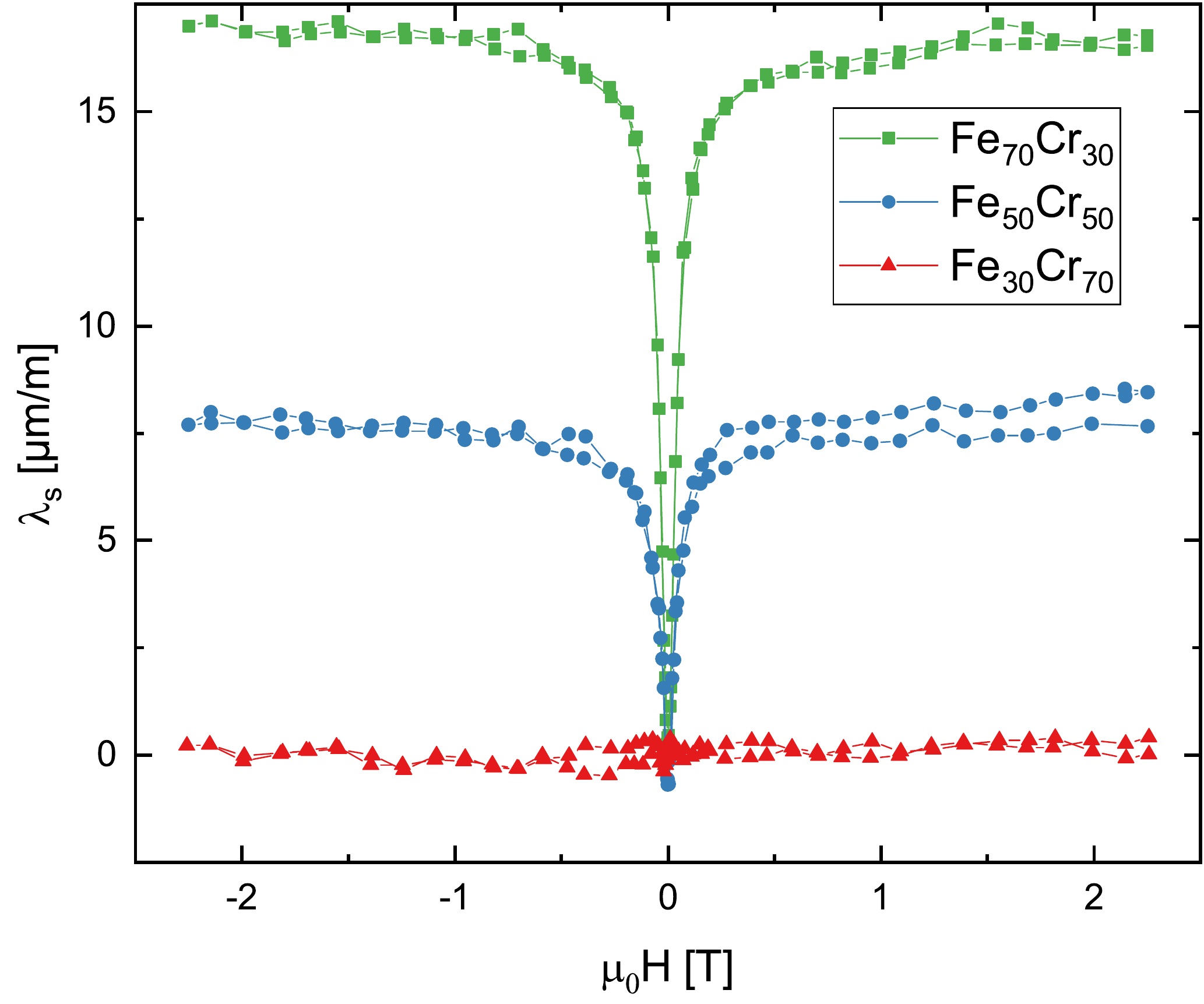}
\caption{Effective magnetostriction as a function of applied magnetic field for three FeCr alloy after HPT-deformation at RT.}
  \label{fig:striction_as-def}
\end{figure}

Assuming a perfectly isotropic material, the magnetostriction measured parallel and perpendicular to the applied magnetic field \cite{dapino2004magnetostrictive}, follows $\lambda_{\parallel}$~=~-2~$\lambda_{\perp}$. The anisotropy of magnetostrictive behaviour is further influenced by the crystal structure and crystallographic directions. Therefore, the effective saturation magnetostriction $\lambda_{s}$ for polycrystalline materials is usually described by 
\begin{equation}\label{eq:lamda}
\lambda_{s} = \frac{2}{3} (\lambda_{\parallel}-\lambda_{\perp})
\end{equation}
According to equation \ref{eq:lamda} the magnetostrictive behaviour of as-deformed FeCr alloys as a function of the applied magnetic field is shown in Figure \ref{fig:striction_as-def}. The saturation value increases with increasing Fe content. 
At a maximal applied field $\upmu_{0} H$ = 2.25~T the signal is close to 0~ppm for the as-deformed Fe$_{30}$Cr$_{70}$ sample, whereas the samples consisting of Fe$_{50}$Cr$_{50}$ and Fe$_{70}$Cr$_{30}$ reach 8~ppm and 17~ppm, respectively. To give a comparison, the magnetostriction of elemental Cr is very small ($\lambda_{s}^{nm}<$ 1~ppm) \cite{lee1969magnetostriction} and elemental Fe exhibits a crystallographic anisotropic magnetostriction, which results in a negative value $\lambda_{s}^{m}$ = -9.3~ppm according to equation \ref{eq:lamda} \cite{dapino2004magnetostrictive,clark1980magnetostrictive}. 
Preliminary SQUID magnetometry measurements conducted on the Fe$_{30}$Cr$_{70}$ sample, exclude a paramagnetic state at room temperature but indicate the presence of superparamagnetic particles, possible existing within the high amount of grain boundaries of the nanocrystalline material \cite{mazilkin2019competition}.
Applying these values and using the $\frac{Fe}{Cr}$ volume fraction \textit{f}, the volumetrically weighted balance between a magnetostrictive $\lambda_{s}^{m}$ and a non-magnetostrictive $\lambda_{s}^{nm}$ component for the effective saturation magnetisation $\lambda_{s}^{eff}$ 
\begin{equation}\label{eq:volume}
\lambda_{s}^{eff} = f \lambda_{s}^{m} + (1 - f) \lambda_{s}^{nm}
\end{equation}
gives deeper insights into the Fe based nanocrystalline alloys \cite{herzer1993nanocrystalline}.
Obviously, this approximation does not hold for our presented alloys as equation \ref{eq:volume} yields an even smaller quantity in $\lambda_{s}$. We can either conclude on an existing solid solution or must assume other mechanisms giving a large positive contribution to the measured magnetostriction.
If we assume separated phases, equation \ref{eq:volume} does not consider contributions of surface spins and grain boundaries. Nevertheless it is known, that both, surface- and bulk- magnetoelastic coupling contribute to the magnetoelastic tensor \cite{szymczak1999mechanisms}, which is further important for nanocrystalline materials \cite{tello2003effective}.
To describe the magnetostrictive behaviour of nanocrystalline materials, Slawska-Waniewska et al. \cite{slawska1996saturation} expanded equation \ref{eq:volume} to 
\begin{equation}\label{eq:slawska}
\lambda_{s}^{eff} = f \lambda_{s}^{m} + (1 - f)\lambda_{s}^{nm}(f) + f \lambda_{s}^{s} \frac{S}{V}
\end{equation}
where $\lambda_{s}^{s}$ describes the interface contribution such as surface spins to the effective saturation magnetostriction. $\frac{S}{V}$ is the surface to volume fraction, which can be rewritten in $\frac{3}{R}$ for spherical grains with an effective grain radius \textit{R}.
For the FeCr system only the third term contribute with a positive value to $\lambda_{s}^{eff}$ keeping in mind, that the measured $\lambda_{s}$ values for Fe$_{50}$Cr$_{50}$ and Fe$_{70}$Cr$_{30}$ presented in Figure \ref{fig:striction_as-def} are positive.
If the second term is neglected and the measured values for $\lambda_{s}^{eff}$ as well as a grain size radius of 50~nm (median grain diameter obtained by TKD measurements - see Figure \ref{fig:TKD}) are used, we can estimate a value for $\lambda_{s}^{s}$.
For the Fe$_{70}$Cr$_{30}$ composition a $\lambda_{s}^{s} \approx$ 400$\cdot$10$^{-6}$~nm is necessary to fit our measurements. Comparing $\lambda_{s}^{s}$ values for FeCr alloys to other Fe based nanocrystalline materials, they differ by one (Fe$_{30}$Cr$_{70}$) or two (Fe$_{50}$Cr$_{50}$ and Fe$_{70}$Cr$_{30}$) orders of magnitude \cite{tello2003effective,slawska1996saturation}.
Therefore, surface spin contributions to the effective saturation magnetostriction are not excluded but cannot be the main contribution of the increased $\lambda_{s}^{eff}$ and consequently, Fe and Cr have to be present in the form of a solid solution.

Hall et al. \cite{hall1960single} used etched single crystals of FeCr and found an increasing magnetostriction ($\lambda_{[100]}$ = 51.7~ppm) when alloying Fe with Cr up to 21.1 at.\% for the crystallographic [100] direction. But they also describe a concurrently decreased value for the [111] direction ($\lambda_{[111]}$ = -2.7~ppm), if compared to elemental Fe \cite{buschow2003physics}. 
According to the relation 
\begin{equation}\label{eq:crystall}
\lambda_{s}^{eff} = \frac{1}{5} (2\lambda_{100} + 3\lambda_{111})
\end{equation}
$\lambda_{s}^{eff}$ results in 19~ppm, which is in excellent agreement with results on Fe$_{70}$Cr$_{30}$, keeping in mind that it is an HPT-deformed sample exhibiting a nanocrystalline microstructure. 
Furthermore, comparable results are found in the work of Bormio-Nunes et al. \cite{bormio2016magnetostriction}. For the identical chemical composition of Fe$_{70}$Cr$_{30}$, they reported $\lambda_{s}$ values between 23~ppm and 47~ppm depending on the measured side of their cubic samples. 
\begin{figure}
\centering
\includegraphics[width=\columnwidth]{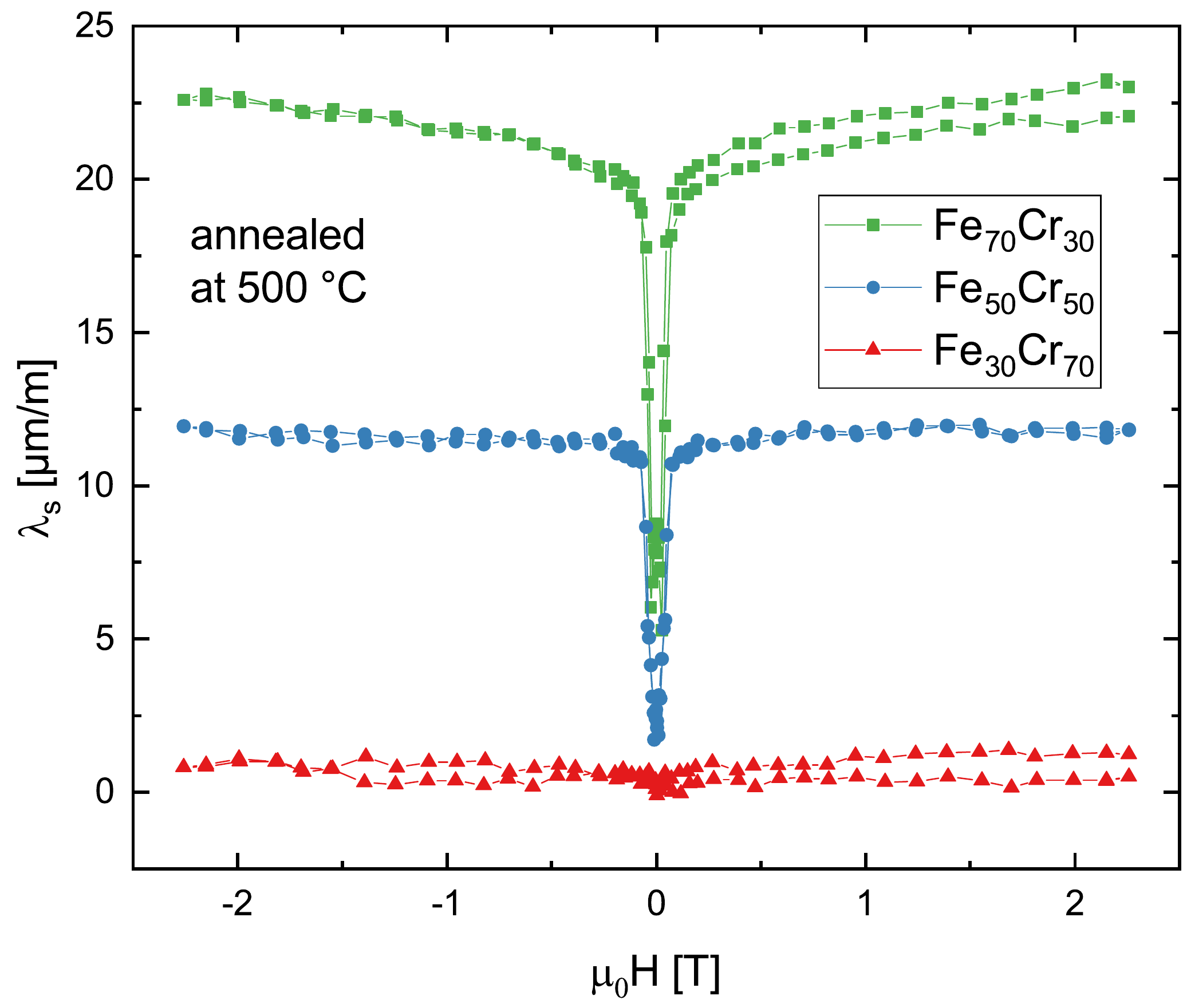}
\caption{Effective magnetostriction as a function of applied magnetic field for three FeCr alloys, HPT deformed at RT and subsequently annealed at 500~$^{\circ}$C for 1~h.}
  \label{fig:striction_annealed}
\end{figure}
The results of magnetostrictive measurements of annealed samples (500~$^{\circ}$C for 1h) are shown in Figure \ref{fig:striction_annealed}.
For all samples, the saturation magnetostriction increases compared to as-deformed samples. The signal for the Fe$_{30}$Cr$_{70}$ sample is slightly positive, the Fe$_{50}$Cr$_{50}$ and Fe$_{70}$Cr$_{30}$ samples reach 12~ppm and 23~ppm, respectively.
The increased magnetostriction with respect to the as-deformed state is thought to originate from a decreased number of crystal defects and a possible reduction of residual stresses \cite{makar1994effects} and further, analysis of EBSD scans (see Figure \ref{fig:TKD}) revealed a slightly stronger texture for the annealed sample, which could also enhance magnetostriction values \cite{han2017tailoring,li2009texture,he2018determination}. Due to the large, positive values, a solid solution is still present for all FeCr compositions but no further improvement of $\lambda_{s}^{eff}$ by an increasing Cr content is found as motivated by Bormio-Nunes et al. \cite{bormio2016magnetostriction}. 

During this work, we successfully produced nanocrystalline FeCr supersaturated solid solutions over a broad chemical composition and clarified that the alloys do not exhibit a two-phase structure in the nanocrystalline regime and confirmed the existence of a solid solution.  
Astonishingly a demixing into separated Fe and Cr phases was suppressed by a remarkable temperature stability of the solid solution up to 520~$^{\circ}$C.
As the principle aim was to decompose the microstructure into a nanocrystalline two-phase material, enabling the formation of an exchange biased material, a more complex annealing treatment and HPT-processing route has to be found, which is part of ongoing studies.
Beforehand, adding substitutional elements could stabilise the microstructure even further and hinder grain growth. This could preserve the nanocrystalline structure during annealing, while a decomposition takes place. 

\section{Conclusion}
Nanocrystalline FeCr alloys consisting of Fe$_{30}$Cr$_{70}$, Fe$_{50}$Cr$_{50}$ and Fe$_{70}$Cr$_{30}$ are processed by HPT-deformation at RT. Upon annealing, the microstructure starts to coarsen between 300~$^{\circ}$C and 500~$^{\circ}$C, which is stronger pronounced for higher Fe contents. All materials investigated exhibit a hardening effect upon annealing with a peak between 300~$^{\circ}$C and 400~$^{\circ}$C followed by a softening at elevated annealing temperatures. In-situ HEXRD annealing experiments point towards a supersaturated solid solution of FeCr. However, concurrent finely dispersed Fe and Cr phases cannot be completely ruled out by microstructural investigations. 
Up to temperatures of 520~$^{\circ}$C, no peak separation or broadening due to phase decomposition is found in HEXRD experiments.
To clarify the ambiguous results from HEXRD experiments in regard of the prevailing phases, magnetostrictive measurements were conducted for HPT-deformed as well as subsequently annealed materials.
Results show a clear deviation with respect to elemental Fe or Cr. The saturation magnetostriction $\lambda_{s}(\upmu_{0}H=2.25~T)$ is 0~ppm (1~ppm), 8~ppm (12~ppm) and 17~ppm (23~ppm) for the as-deformed (500~$^{\circ}$C annealed) alloys consisting of Fe$_{30}$Cr$_{70}$, Fe$_{50}$Cr$_{50}$ and Fe$_{70}$Cr$_{30}$, respectively. After excluding a magnetostrictive contribution from interfaces, these values are explained by the formation of a supersaturated solid solution by HPT-deformation, which persists during annealing and validates findings from the synchrotron HEXRD measurements.
To decompose these FeCr solid solutions and to enable exchange bias effects, more complex processing routes are necessary.

\textbf{Declaration of Competing Interest}
The authors declare, that they have no known competing financial interests or personal relationships, that could have appeared to influence the work reported in this paper.

\textbf{Acknowledgements}
The synchrotron measurements leading to these results have been performed at beamline P21.2 at PETRA III under proposal I-20190577. 
The authors thank DESY (Hamburg, Germany), a member of the Helmholtz Association HGF, for the provision of experimental facilities and gratefully acknowledge the assistance by Timo Müller who provided a lot of help and commitment to our experiments. They further thank F. Spieckermann and C. Gammer for their help with XRD-data processing.
This project has received funding from the European Research Council (ERC) under the European Union’s Horizon 2020 research and innovation programme (Grant No. 757333).

  \bibliography{arxiv_mstric_FeCr}

\newpage
\thispagestyle{empty}
\begin{figure*}
\centering
\renewcommand\figurename{Supplementary}
 \setcounter{figure}{0}
\includegraphics[width=\textwidth]{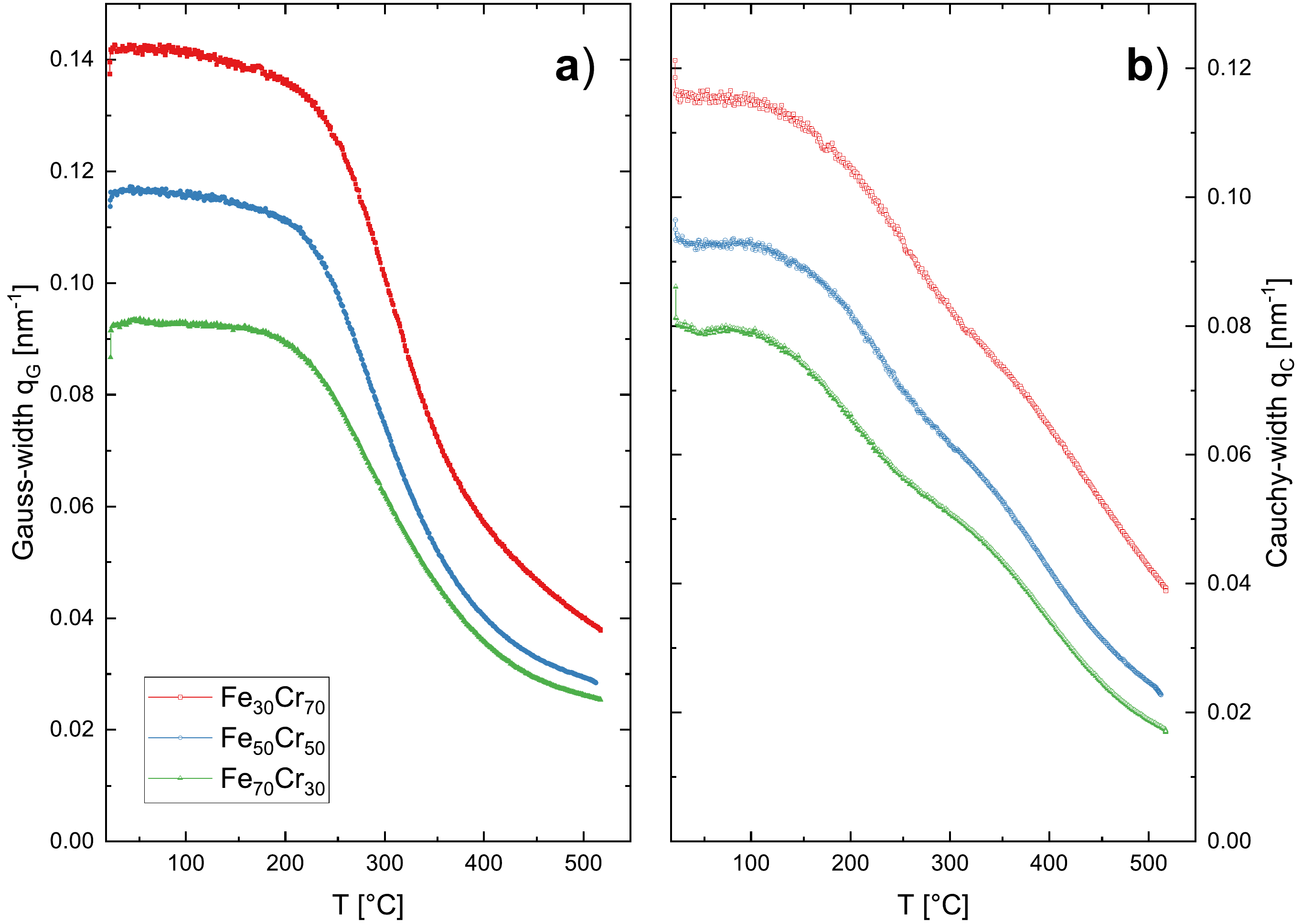}
\caption{The [110] peak of the in-situ synchrotron annealing experiment is fitted by a single Voigt-peak. The FWHM is separated into a Gaussian \textbf{a)} and Cauchy \textbf{b)} contribution and plotted against the increasing temperature during heating.}
  \label{fig:DESY_voigt}
\end{figure*}

\end{document}